\documentclass{article}
\usepackage{fleqn}
\usepackage{amsmath,amssymb}

\begin{document}
\setcounter{section}{5}
\setcounter{equation}{112}
\begin{center}
{\bf\Large Cosmological evolution  of the cosmological plasma with interpartial scalar interaction.\\[12pt]  III. Model with the attraction of the like scalar charged particles.} \\[12pt]
Yu.G. Ignat'ev\\
N.I. Lobachevsky Institute of Mathematics and Mechanics, Kazan Federal University, \\ Kremleovskaya str., 35, Kazan, 420008, Russia
\end{center}

\begin{abstract} On the basis of the relativistic kinetic theory the mathematical model of cosmological plasmas with an attraction of the like charged scalar particles is formulated. It is shown, that cosmological the model, based on a classical scalar field with an attraction, is unsatisfactory, that leads to necessity of attraction of phantom models of a scalar field for systems with an attraction.
\end{abstract}

\section{Features of statistical systems of particles with an attraction of like scalar charges}

\noindent In the second part of article we have entered a Lagrange function for a scalar massive field (74)\footnote{ Numbering of formulas by all parts of article through - all formulas to (34) inclusive concern the first part of article [1], all formulas from (35) to (102) inclusive concern the second part [2].}

\begin{equation} \label{GrindEQ__74_}
L_{s} =\frac{\varepsilon }{8\pi } \left[g^{ik} \Phi _{,i} \Phi _{,k} +\left(\frac{R}{6} -m_{s}^{2} \right)\Phi ^{2} \right]
\end{equation}
in which to value $\varepsilon =+1$ there corresponds a case of mutual pushing away the same scalar charged particles, and to value $\varepsilon =-1$ corresponds a mutual attraction the same scalar charged particles For more exact finding-out of physical sense of the parametres model statistical system with scalar interaction we will consider the system consisting of one particle with certain values of phase co-ordinates. Let we choose a frame of reference, in which four-dimensional space representable in the form of direct product three-dimensional riemann spaces  $V_{3} $ of time T: $V_{4} =V_{3} \times T$. According to [3] invariant function of distribution of such system looks like\footnote{ We remind, that we accept universal system of units:  $G=c=\hbar =1$ .}:

\begin{equation} \label{GrindEQ__103_}
F(x,P)=(2\pi )^{3} D(x|x_{0} )\delta ^{3} (P^{\alpha } -P_{0}^{\alpha } ),
\end{equation}
where $D(x|x_{0} )$ - the invariant symmetric point-to-point Dirac function connected with scalar density $\Delta (x_{1} |x_{2} )$, usually and named $\delta $-function of Dirac, a relation:

\begin{equation} \label{GrindEQ__104_}
\Delta (x_{1} |x_{2} )=\frac{1}{\sqrt{-g} } D(x_{1} |x_{2} ) .
\end{equation}
The invariant Dirac function determinate in an ordinary way:

\[\int _{X_{2} }D(x_{1} |x_{2} ) f(x_{2} )dX_{2} =\left\{\begin{array}{c} {f(x_{1} ),{\rm \; }x_{1} \in X_{2} ;} \\ {0,{\rm \; \; \; \; \; \; \; }x_{1} \notin X_{2} } \end{array}\right. ,\]
where f(x) is arbitrary tensore field in $V_{n} $ and  $dX=\sqrt{-g} dx^{1} \cdot dx^{2} \cdot \cdot \cdot dx^{n} $  is the invariant volume element of $V_{n} $. Thus, the vector of density of a stream of number of particles (46) for a particle with certain phase co-ordinates becomes:

\begin{equation} \label{GrindEQ__105_}
n^{i} =D(x|x_{0} )\iiint \nolimits _{P(x)}\frac{d^{3} P}{\sqrt{-g} P_{4}^{0} } \delta ^{3} (P^{\alpha } -P_{0}^{\alpha } ) \equiv D(x|x_{0} )\frac{P_{0}^{i} }{P_{4}^{0} } ,
\end{equation}
where $P_{4}^{0} $ is a positive root of the mass surface equation (8):

\[(P,P)=m_{*}^{2} .\]
Calculating now full number of particles in configuration space, we will receive:
\begin{equation} \label{GrindEQ__106_}
N(t)=\iiint \nolimits _{V_{3} }n^{i}  u_{i} dV_{3} =\iiint \nolimits _{V_{3} }\sqrt{-g}  D(x|x_{0} )d^{3} x=1,
\end{equation}
Where $u_{i} $ - unit timelike a normal to a hypersurface (a vector of speed of the observer). Thus, \eqref{GrindEQ__103_} defines correct normalization of distribution functions of a single particle with the fixed phase co-ordinates. Calculating by means of \eqref{GrindEQ__103_} and (48) trace of an EMT of particles, we will receive:

\begin{equation} \label{GrindEQ__107_}
T_{p} =\frac{m_{*}^{2} }{P_{4}^{0} } D(x|x_{0} )
\end{equation}
And, thus, according to (60) we will find scalar density of a single charge:

\begin{equation} \label{GrindEQ__108_}
\sigma =q\frac{m_{0} +q\Phi }{P_{4}^{0} } D(x|x_{0} ).
\end{equation}
In particular, in a case motionless in space Minkovsky particles we will receive from here:

\begin{equation} \label{GrindEQ__109_}
\sigma =qsgn(m_{0} +q\Phi )\delta ^{3} (r).
\end{equation}

\noindent In case of nonzero rest mass of a particle and a zero scalar field the classical result for newton charge density is received from here.

Let's consider the equations of static central-symmetric and conformal noninvariant scalar field                                $\square \Phi +m_{s}^{2} \Phi =-4\pi \varepsilon \sigma $                                                                                               (85) in space Minkovsky with density of a scalar charge \eqref{GrindEQ__109_}:

\begin{equation} \label{GrindEQ__110_}
\frac{1}{r^{2} } \frac{d}{dr} \left(r^{2} \frac{d}{dr} \Phi \right)-m_{s}^{2} \Phi =4\pi \varepsilon q\ {\rm sgn}(m_{0} +q\Phi )\delta ^{3} (r).
\end{equation}
Considering a known relation:\footnote{ See, for example, [4].}

\[\Delta \frac{1}{r} \equiv \frac{1}{r^{2} } \frac{d}{dr} \left(r^{2} \frac{d}{dr} \frac{1}{r} \right)=-4\pi \delta ^{3} (r),\]
Let's find formal, convergent on infinity the solution of the equation \eqref{GrindEQ__110_}:

\noindent

\begin{equation} \label{GrindEQ__111_}
\Phi =\varepsilon \ {\rm sgn}(m_{0} +q\Phi )_{r\to 0} \frac{qe^{-m_{s} r} }{r}
\end{equation}

\noindent Further, as $\Phi (0)\to \infty $ sign function is defined it is only by sign expressions $q\Phi (r)$ at $r\to 0$:

\begin{equation} \label{GrindEQ__112_}
\Phi =\varepsilon \ {\rm sgn}(q\Phi )_{r\to 0} \frac{qe^{-m_{s} r} }{r}
\end{equation}
Multiplying both parts of the solution \eqref{GrindEQ__111_} on q, we will come to conclusion, that the sign on function $\varphi =q\Phi $ is defined is only by sign the factor $\varepsilon $:

\begin{equation} \label{GrindEQ__113_}
sgn(q\Phi )=\varepsilon
\end{equation}
Is very important property of a classical scalar field. Just this property also shows, that to values $\varepsilon =-1$ there corresponds negative potential energy   $U=q\Phi $, i.e., an attraction, and to case $\varepsilon =+1$ - positive potential energy, i.e., pushing away. Thus, in both cases we have potential of interaction \eqref{GrindEQ__112_} types of potential of the Yukawa. We will notice, that a classical case of an attraction ($\varepsilon =-1$) for a massless scalar field, actually, corresponds, to so-called, phantom fields. In this scheme at nonzero mass of a scalar field its potential submits to the law of the Yukawa. In this article we also will consider also phantom scalar fields with a negative massive member - more precisely speaking we will keep a sign on a massive member in a Lagrange function, but we will change a sign on a "kinetic" member\footnote{ For the conformal noninvariant scalar field it is necessary to lower a member with scalar curvature in the ratio (114).}:

\begin{equation} \label{GrindEQ__114_}
L_{s} =-\frac{1}{8\pi } \left[g^{ik} \Phi _{,i} \Phi _{,k} +\left(\frac{R}{6} +m_{s}^{2} \right)\Phi ^{2} \right].
\end{equation}

With an attraction of the same charged particles we will explain the reason of introduction of such Lagrange function for statistical systems later. The energy-momentum tensor of a scalar field concerning function of a Lagrange function \eqref{GrindEQ__114_} looks like:

\noindent

\begin{equation} \label{GrindEQ__115_}
T_{s}^{ik} =\frac{1}{8\pi } \left(-\frac{4}{3} \Phi ^{i} \Phi ^{k} +\frac{1}{3} g^{ik} \Phi _{,j} \Phi ^{,j} +m_{s}^{2} g^{ik} \Phi ^{2} \right).
\end{equation}
Thus definition of scalar density of charges $\sigma $ (58 [2] remains invariable:

\begin{equation} \label{GrindEQ__116_}
\sigma =\sum \frac{d\ln |m_{*} |}{d\Phi }  T_{p}
\end{equation}
and the equation of a scalar field becomes:

\begin{equation} \label{GrindEQ__117_}
\square \Phi -m_{s}^{2} \Phi -\frac{R}{6} \Phi =4\pi \sigma
\end{equation}
Solving similar previous the problem for a central-symmetric static field in space Minkovsky, we will receive:

\begin{equation} \label{GrindEQ__118_}
\Phi = {\rm sgn}(m_{0} +q\Phi )_{r\to 0} \frac{q\sin m_{s} r}{r} .
\end{equation}
As in this case the potential in the beginning of co-ordinates is limited, calculating a limit, we will receive from here:

\[\Phi (0)=qm_{s} {\rm sgn}(m_{0} +q\Phi (0)).                                                                                             \]
It is easy to show, that this algebraic equation has only one root:

\noindent

\begin{equation} \label{GrindEQ__119_}
\Phi (0)=qm_{s} .
\end{equation}
Thus, despite seeming formal complexity of the found solution, we will receive from \eqref{GrindEQ__118_} and \eqref{GrindEQ__119_} simple solution of the equation for a scalar field of a single based scalar charge:

\begin{equation} \label{GrindEQ__120_}
\Phi =q\frac{\sin m_{s} r}{r} .
\end{equation}
Hence, massive phantom scalar fields are nonsingular in sources and far-ranging fields with an attraction of the same charged particles.

\section{Self-consistent cosmological model for locally equilibrium plasma with an attraction the same scalar charged particles}

\textbf{}

Calculating an energy-momentum tensor of a scalar field concerning the metrics of spatially-flat Freedman Universe                  
\begin{equation}\label{GrindEQ__121_}
ds^{2} =dt^{2} -a^{2} (t)(dx^{2} +dy^{2} +dz^{2} ),
\end{equation}
 let's receive from \eqref{GrindEQ__115_} density of energy and pressure of a scalar field with an attraction the same scalar charged particles:

\begin{equation} \label{GrindEQ__122_}
E_{s} =\frac{1}{8\pi } (-\dot{\Phi }^{2} \mp m_{s}^{2} \Phi ^{2} );\quad P_{s} =-\frac{1}{8\pi } (\frac{1}{3} \dot{\Phi }^{2} \mp m_{s}^{2} \Phi ^{2} ),
\end{equation}
Where the top sign corresponds to a classical scalar field with an attraction, the bottom sign - to a phantom scalar field with an attraction \footnote{ Apparently, the attraction-pushing away factor the same scalar charged particles is completely defined by a sign on a kinetic member in a Lagrange function, therefore further we for ü to a phantom field will not add «with an attraction».}. Thus, the density of energy of a classical scalar field with an attraction is strictly negative, as interferes with use of a classical scalar field with potential of type of the Yukawa as model of physical vacuum. In a case of a phantom scalar field with the massive member, corresponding to imaginary mass, such obstacle does not arise. This factor, apparently, also is defining correct relativistic model of scalar interaction of particles.

In particular, the class of cosmological models with an inflationary initial stage for scalar fields with an attraction unequivocally leads to a choice of phantom scalar fields as Einstein's equations for a classical scalar field with an attraction look like $\dot{a}^{2} =-b^{2} $, i.e., have no valid solutions. Let's notice, that the trace of an energy-momentum tensor of a scalar field is equal:

\begin{equation} \label{GrindEQ__123_}
T_{s} =E_{s} -3P_{s} =\frac{1}{2\pi } m_{s}^{2} \Phi ^{2} ,
\end{equation}
 and too:

\begin{equation} \label{GrindEQ__124_}
E_{s} +P_{s} =-\frac{\dot{\Phi }^{2} }{4\pi } .
\end{equation}
The equation for a scalar field becomes:

\begin{equation} \label{GrindEQ__125_}
\ddot{\Phi }+3\frac{\dot{a}}{a} \dot{\Phi }\pm m_{s}^{2} \Phi =4\pi \sigma ,
\end{equation}
where besides the top sign corresponds to a classical scalar field with an attraction, the bottom sign - to a phantom scalar field.

\noindent The equilibrium density of a scalar charge, $\sigma $, in the equation \eqref{GrindEQ__125_} is described by the formula \eqref{GrindEQ__73_}:

\begin{equation} \label{GrindEQ__126_}
\sigma =\sum _{a} \frac{2S+1}{2\pi ^{2} } q(m+q\Phi )^{3} \int _{0}^{\infty } \frac{\sinh^{2} xdx}{e^{-\gamma _{a} +\lambda _{*} \cosh x} \pm 1} ,
\end{equation}
in which we have put a mass function equal$m_{*} =|m+q\Phi |$, and $\lambda _{*} =m_{*} /\theta $. If we make the additional assumption of preservation of a scalar charge in all reactions then we will receive a condition on chemical potential of scalar charged particles:

\begin{equation} \label{GrindEQ__127_}
a^{3} (t)\sum q_{a} n_{a}  =a^{3} (t)\sum q_{a} \frac{2S+1}{2\pi ^{2} } m_{*}^{3} \int _{0}^{\infty } \frac{\sinh^{2} x\cosh xdx}{e^{-\gamma _{a} +\lambda _{*} \cosh x} \pm 1}  =Const,
\end{equation}
where summation is spent on charged particles all scalar. In particular, if to assume, that particles of one grade «a» with opposite scalar charges can àííèãèëèðîâàòü:

\begin{equation} \label{GrindEQ__128_}
a+\bar{a}=\gamma +\gamma ,
\end{equation}
from conditions of chemical balance (67) we will receive communication of reduced chemical potentials of scalar charged particles and antiparticles:

\begin{equation} \label{GrindEQ__129_}
\bar{\gamma }_{a} =-\gamma _{a} .
\end{equation}
Thus taking into account the algebraic equation \eqref{GrindEQ__127_} we receive the closed system of the equations for cosmological model (99), (101), \eqref{GrindEQ__125_}.

\noindent E-mail: ignatev\_yu@rambler.ru

\end{document}